\documentclass[aps,prb,twocolumn,floatfix,floats,superscriptaddress,showpacs]{revtex4}
\usepackage[english]{babel}
\usepackage[T1]{fontenc}
\usepackage{dcolumn}
\usepackage{amsmath}
\usepackage{graphicx}

\def\bra#1{\left\langle #1 \right|}             
\def\ket#1{\left| #1 \right\rangle}             

\def\be{\begin{equation}}
\def\ee{\end{equation}}
\def\bea{\begin{eqnarray}}
\def\eea{\end{eqnarray}}
\newcommand{\rb}{\mathbf{r}}

\begin{document}

\title{Coherent electronic transport in a multimode quantum channel \\
       with Gaussian-type scatterers}

\author{Jens Hjorleifur Bardarson}
\affiliation{Science Institute, University of Iceland, 
            Dunhaga 3, IS-107 Reykjavik, Iceland}
\author{Ingibjorg Magnusdottir}
\affiliation{Science Institute, University of Iceland, 
            Dunhaga 3, IS-107 Reykjavik, Iceland}
\author{Gudny Gudmundsdottir}
\affiliation{Science Institute, University of Iceland, 
            Dunhaga 3, IS-107 Reykjavik, Iceland}
\author{Chi-Shung Tang}
\affiliation{Physics Division, National Center for Theoretical Sciences, 
            P.O.~Box 2-131, Hsinchu 30013, Taiwan}
\author{Andrei Manolescu}
\affiliation{Science Institute, University of Iceland, 
        Dunhaga 3, IS-107 Reykjavik, Iceland}
\author{Vidar Gudmundsson}
\affiliation{Science Institute, University of Iceland, 
        Dunhaga 3, IS-107 Reykjavik, Iceland}
\date{\today}

\begin{abstract}
Coherent electron transport through a quantum channel in the presence 
of a general extended scattering potential is investigated using a 
$T$-matrix Lippmann-Schwinger approach. The formalism is applied to a quantum wire with 
Gaussian type scattering potentials, which can be used to model a single 
impurity, a quantum dot or more complicated structures in the wire. 
The well known dips in the conductance in the presence of attractive
impurities is reproduced. A resonant transmission peak in the conductance 
is seen as the energy of the incident electron coincides with an energy level 
in the quantum dot. The conductance through a quantum wire in the presence of 
an asymmetric potential are also shown. In the case of a narrow potential 
parallel to the wire we find that two dips appear in the same subband which 
we ascribe to two quasi bound states originating from the next evanescent mode.
\end{abstract}

\maketitle

\section{\label{sec:Intro}Introduction}
In studies of electronic transport in mesoscopic or nanoscale
quasi-one-dimensional systems it has strikingly been found that the
conductance manifests
quantization~\cite{WHB+88,wha88,wee88b,Houten89} when the
electronic phase-coherent length $l_{\phi}$ is greater than the
system size.  The transport properties of these
systems have been successfully described within the
Landauer-B\"uttiker framework.~\cite{Landauer1,Landauer2,Landauer3,BILP85,Buttiker2, Buttiker3}  
Later on, the influence of a single impurity on the conduction in quantum
channels has attracted a great deal of attention since the impurity
inside or near the conducting channel may destroy the conduction
quantization, as has been demonstrated theoretically
~\cite{chu89,Bagwell90,tek91,nix91,LLS92,tak92,kun92,chu94,boe00} and
experimentally.~\cite{fai90}  

The impurities are usually assumed to be zero-range, i.e.\ of a delta-function type, in theoretical
considerations.\cite{Bagwell90,VV2002,CM03}
However, a real physical impurity must be of finite range and therefore
modeled by an extended potential. 
In this paper we discuss coherent transport properties of quantum channels in
the presence of extended scattering potentials. If the
potentials are very extended they can describe the effects of a tunable central gate.\cite{soh01}

The method for the coherent electron transport through the quantum 
channel employs the Lippmann-Schwinger (LS) formalism. Its relation to the
$T$-matrix and the Landauer formula in terms of elements of the scattering
matrix is discussed in Sec.~\ref{sec:Theory}.
The conductance of a quantum channel in the presence of extended Gaussian type scattering
potentials obtained by numerical calculations is given in
Sec.~\ref{sec:Results}. This section consists of three parts. First we discuss the
scattering by a single Gaussian potential. The physics of the transport is
examined more closely by visualizing the most important processes by help of
the LS equation. In the second part we model a quantum
dot embedded in the wire by a combination of two Gaussian potentials, and use
that to obtain the conductance of such a setup. The effect of the shape of an
extended scattering potential on the conductance is
considered in the last part. In Sec.~\ref{sec:summary} we summarize and
discuss the main results of the paper.

\section{\label{sec:Theory}Scattering by a Potential}
\begin{figure}[tbh]
      \includegraphics[width=0.35\textwidth]{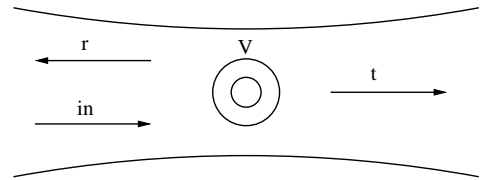}
      \caption{Schematic view of the system. An incoming wave is partly 
               transmitted and partly reflected by
               the finite range scattering potential $V$.}
\label{fig:Schematic}
\end{figure}

We consider a quantum wire connected adiabatically to reservoirs as
schematically shown in Fig.~\ref{fig:Schematic}.  
The uniformity of the wire is broken by a scattering potential of finite
extent in the wire.
The wave function of a single-electron state with energy $E$, in the
quasi-one-dimensional quantum wire is described by the Schr{\"o}dinger
equation 
\begin{equation}
\label{eq:SchrodingerFull}
      \left(-\frac{\hbar^2}{2m}\left[
      \frac{\partial^2}{\partial x^2} 
      + \frac{\partial^2}{\partial z^2} \right] +
      V_c(x) + V(\mathbf{r})\right)\psi_E(\mathbf{r}) =  E\psi_E(\mathbf{r}).
\end{equation}
The electron is confined in the $x$-direction by the confinement potential
$V_c(x)$ but is free to propagate in the $z$-direction. $V(\mathbf{r})$ is a
finite range scattering potential.  
The transverse modes
\begin{equation}
\label{eq:TransverseModes}
      \left(-\frac{\hbar^2}{2m}\frac{d^2}{dx^2} + V_c(x) \right)\chi_n(x) = \varepsilon_n\chi_n(x),
\end{equation}
are assumed to be known.

\begin{figure}[tbh]
      \includegraphics[width=0.35\textwidth]{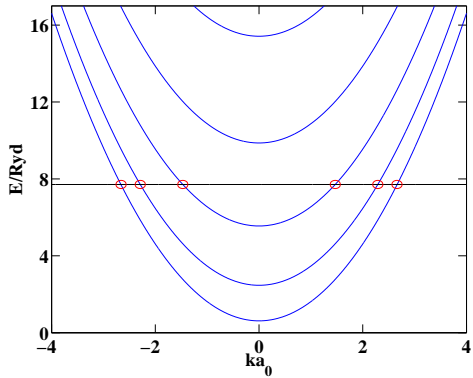}
      \caption{(Color online). The energy levels of a quantum wire with a hard wall confinement. 
               For a given energy $E$ (horizontal line) only a finite number of
               propagating modes are active (red markers), but higher lying evanescent modes
               can serve as intermediate states. This specific case is for a
               wire of width $L = 4a_0$. Energy is scaled
               in Rydbergs (Ryd) and lengths in Bohr radii ($a_0$), cf.\ Sec.~\ref{sec:Results}.}
\label{fig:EnergyBands}
\end{figure}

Outside the range of the scattering potential an electron with energy $E$ is
in a linear combination of the eigenfunctions
of $H_0 = H - V(\mathbf{r})$. These eigenfunctions are referred to as modes and can be written
\begin{equation}
\label{eq:Modes}
      \phi_{nE}^\pm(\mathbf{r}) = \frac{1}{\sqrt{k_n(E)}}e^{\pm ik_n(E)z}\chi_n(x),
\end{equation}
where $E = \hbar^2k_n^2(E)/2m + \varepsilon_n$ and the modes are
normalized to carry unit probability
current.\cite{fisher81:6851,BruusFlensberg} The $+$ (-) refers to waves incident from
the left (right).
If $\varepsilon_n < E$, $k_n(E)$ is real and $\phi_{nE}^\pm$ are
propagating waves (see Fig.\ \ref{fig:EnergyBands}). If on the other hand 
$\varepsilon_n > E$, $k_n(E)$ is purely
imaginary and the eigenfunctions must be exponentially decaying, evanescent modes. 

Scattering states are special solutions of Eq.~\eqref{eq:SchrodingerFull},
where e.g.\ an electron in propagating mode $n$ impinges on the scattering 
potential from the left and is backscattered or transmitted into other modes. 
In the far zone, outside the range of the scattering potential, the scattering states have the
boundary condition
\begin{equation}
\label{eq:ScatteringState+}
      \psi_{nE}^+(\mathbf{r}) = 
  \begin{cases}
      \phi_{nE}^{+}(\mathbf{r}) + \sum_{m,\text{prop}}
      r_{mn}\phi_{mE}^{-}(\mathbf{r}), & z \rightarrow -\infty, \\
      \sum_{m,\text{prop}} t_{mn}\phi_{mE}^{+}(\mathbf{r}),  & z \rightarrow \infty.
  \end{cases}
\end{equation}
$t_{mn}$ is the probability amplitude for a particle in mode $n$ in the left
lead to scatter out in mode $m$ in the right lead. $r_{mn}$ gives
the probability amplitude for reflection into mode $m$. By
restricting all the modes to carry unit current the transmission amplitudes
coincide with the elements of the scattering matrix.\cite{BILP85}
Hence, the zero temperature linear response conductance of the system can be
obtained by the Landauer formula~\cite{BILP85,fisher81:6851}
\begin{equation}
\label{eq:Landauer}
      G  = \frac{2e^2}{h}\text{Tr}[t^\dagger t],
\end{equation}
where the matrix of the transmission amplitudes $t$ is evaluated at the Fermi energy.

\subsection{Scattering States and the Lippmann-Schwinger Equation}
We expand the scattering states in the transverse modes
\begin{equation}
\label{eq:SSexpansion}
      \psi_{nE}^+(\mathbf{r}) = \sum_m \varphi_{mE}^n(z)\chi_m(x),
\end{equation}
where the $n$ denotes the number of the incident mode (cf.\
Eq.~\eqref{eq:ScatteringState+}). Introducing this
expansion in the Schr{\"o}dinger equation~\eqref{eq:SchrodingerFull}, 
multiplying with $\chi_{m'}^*(x)$, integrating over $x$ and letting 
$m \leftrightarrow m'$, one obtains a coupled mode 
equation\cite{Bagwell90,CM03} 
\begin{equation}
\label{eq:SchrodingerZ}
      \left( \frac{d^2}{dz^2} + k_m^2(E) \right)\varphi_{mE}^n(z) =
      \frac{2m}{\hbar^2}\sum_{m'}V_{mm'}(z)\varphi_{m'E}^n(z),
\end{equation}
where
\begin{equation}
\label{eq:Vmm'}
      V_{mm'}(z) = \int dx \, \chi_m^*(x) V(\mathbf{r})
      \chi_{m'}(x).
\end{equation}
Defining a mode Green's function as
\begin{equation}
\label{eq:Green_nE}
      \left( \frac{d^2}{dz^2} + k_n^2(E) \right)\mathcal{G}_{nE}^0(z,z') = \delta(z-z'),
\end{equation}
the solution to Eq.~\eqref{eq:SchrodingerZ} can be written in the form of an 
effective 1D LS equation
\begin{equation}
\label{eq:LSZsol}
  \begin{split}
        \varphi_{mE}^n(z) &= \varphi_{mE}^{n0}(z) + \frac{2m}{\hbar^2}\sum_{m'}\int
        dz'\, \mathcal{G}_{mE}^0(z,z')\\ &\times V_{mm'}(z')\varphi_{m'E}^n(z'),
  \end{split}
\end{equation}
where $\varphi_{mE}^{n0}(z) = \delta_{nm}\exp(ik_m(E)z)/\sqrt{k_m(E)}$. 
Note that the solutions of the LS equation~\eqref{eq:LSZsol} obey the same
normalization as the incident mode.
Inserting the explicit form of the Green's function, 
\begin{equation}
\label{eq:Gn}
  \mathcal{G}_{nE}^0(z,z') = -\frac{i}{2k_n(E)}e^{ik_n(E)|z-z'|},
\end{equation}
in Eq.~\eqref{eq:LSZsol} and taking the limit 
$z\rightarrow \infty$, one obtains by comparison with
Eq.~\eqref{eq:ScatteringState+} the probability amplitudes for
forward scattering
\begin{equation}
\label{eq:t_mn}
  \begin{split}
        t_{mn} &= \delta_{mn} + \frac{m}{i\hbar^2}\sum_{m'}\int dz'\,
        \frac{1}{\sqrt{k_m(E)}}e^{-ik_m(E)z'} \\ &\times V_{mm'}(z')\varphi_{m'}^n(z').
  \end{split}
\end{equation}

\subsection{\label{ch:ScattTmatrix}Scattering States and the $T$-matrix}
Having obtained the transmission amplitudes in terms of the configuration
space wave function we now find them in terms of matrix elements of a 
transition operator. Inserting the definition~\eqref{eq:Vmm'} of $V_{mm'}$ into
relation~\eqref{eq:t_mn} and using the expansion~\eqref{eq:SSexpansion} one
obtains
\begin{equation}
\label{eq:t_nm2}
      t_{mn} = \delta_{mn} + \frac{m}{i\hbar^2}\langle mk_m|\hat{V}|\psi_{nE}^+\rangle,
\end{equation}
where in general $\langle \mathbf{r} | nq\rangle = (\exp(iqz)/\sqrt{|q|})\chi_m(x)$.
Similarly, using the LS equation~\eqref{eq:LSZsol} in the same expansion, we see that
\begin{equation}
\begin{split}
      \psi_{nE}^+(\mathbf{r}) = \phi_{nE}^+(\mathbf{r}) +\int d^3r'\, 
      \mathcal{G}_0(\mathbf{r},\mathbf{r}';E)V(\mathbf{r}')\psi_{nE}^+(\mathbf{r}')
\end{split}
\end{equation}
where
\begin{equation}
\label{eq:G03D}
      \mathcal{G}_0(\mathbf{r},\mathbf{r}';E) = \frac{2m}{\hbar^2}\sum_m
      \chi_{m}^*(x)\mathcal{G}_{mE}^0(z,z')\chi_{m}(x).
\end{equation}
This is just the conventional LS equation and therefore by defining the
$T$-matrix by the relation
\begin{equation}
\label{eq:Tmatrix3D}
      \langle m k' | \hat{T} | n k_n\rangle = \langle m k' |\hat{V} | \psi_{nE}^+ \rangle,
\end{equation}
it satisfies the operator LS equation
\begin{equation}
\label{eq:TopLS}
      \hat{T} = \hat{V} + \hat{V}\hat{\mathcal{G}}_0(E)\hat{T}.
\end{equation}
In the eigenfunction basis of $\hat{H}_0$ the LS equation (\ref{eq:Tmatrix3D}) 
is transformed into\cite{CM03}
\begin{equation}
\label{eq:TopLSeig2}
\begin{split}
      T_{mn}(k,k_n) &= V_{mn}(k,k_n) + \frac{m}{\pi\hbar^2}\sum_l\int dq\, |q|
      \\ &\times \frac{V_{ml}(k,q)T_{ln}(q,k_n)}{k_l^2-q^2 + i\eta},
\end{split}
\end{equation}
where the notation $V_{ml}(k,q) = \bra{mk}\hat{V}\ket{lq}$ has been
introduced. We have used that in our current normalization the unity operator is
\begin{equation}
\label{eq:UnityOperator}
      \hat{1} = \sum_l\int dq\, \frac{|q|}{2\pi} \ket{lq}\bra{lq},
\end{equation}
and that the Green's function is
\begin{equation}
\label{eq:Green_l(k)}
      \bra{lq}\hat{\mathcal{G}}_0(E)\ket{l'q'} =
      \frac{2\pi}{|q|}\frac{\delta_{ll'}\delta(q-q')}
      {E-\frac{\hbar^2q^2}{2m} - \varepsilon_l+i\eta}. 
\end{equation}
The numerical solution of Eq.~\eqref{eq:TopLSeig2} is briefly discussed in
App.~\ref{sec:appendix}.

In the above notation the transmission amplitudes can be written
\begin{equation}
\label{eq:t_nm3}
      t_{mn} = \delta_{mn} + \frac{m}{i\hbar^2}T_{mn}(k_m,k_n).
\end{equation}
Note that through Eq.~\eqref{eq:TopLSeig2} $T_{mn}(k_m,k_n)$ depends on
$T$-matrix elements both on and off the
energy shell ($\hbar^2k^2/2m + \varepsilon_n$ is equal to $E$ on the energy shell but
not off it).

We have thus managed to link the elements of the transmission matrix to
matrix elements of the transition operator $\hat{T}$. The result is
intuitively appealing, transmission from one channel to another is related to the transition from an
eigenstate in one channel to an eigenstate in another. 
The formalism is quite general with respect to
the type of wire confinement and scattering potential of finite range.
A change of either one requires only a new evaluation of matrix elements
of the scattering potential and the energy spectrum of $H_0$.

\section{\label{sec:Results}Results}
In our model we use two potentials, one describing the
confinement and the other representing the scattering potential.
We use two different models for either one of these. 
The confining potential, characterizing the extent of 
the quantum channel, can be described by either a hard wall potential 
\begin{equation}
  \label{eq:VcHardWall}
  V_{c,\text{hard}}(x)= 
  \begin{cases}
        0 & 0 < x < L \\
        \infty & \text{elsewhere},
  \end{cases}
\end{equation}
with eigenvalues $\varepsilon_n = \hbar^2\pi^2(n/L)^2/2m$, $n=1,2,\ldots$,
or parabolic walls
\begin{equation}
  \label{eq:VcParabolic}
      V_{c,\text{parabolic}}(x) = \frac{1}{2}m\omega^2x^2r, 
\end{equation}
with eigenvalues $\varepsilon_n =  (n + 1/2)\hbar\omega$, $n=0,1,\ldots$. Note
that the first mode in the hard wall
confinement has $n=1$ but $n=0$ in the parabolic case. 

By a proper choice of model for the scattering potential one can study a
multitude of different systems. In the following subsections we will by use of 
Gaussian functions model a single impurity by a single Gaussian, a
quantum dot embedded in the wire by a sum of two Gaussians of different size
and finally study shape effects of the scattering
potential by asymmetric Gaussians. Even though we have chosen to use only 
Gaussian potentials we stress that the formalism is general and can be used
with more complicated potentials. By cleverly choosing the functional form of the
scattering potential it can be adapted to describe many experimental setups. 

In this paper we scale all energies in Rydbergs (Ryd) and lengths in units of
Bohr radii ($a_0$). The calculations are independent of exact material
constants but it can be useful to keep in mind that in GaAs $a_0 = 9.79$ nm and Ryd $=5.93$
meV.

\subsection{Single Gaussian Potential}
A single Gaussian potential
\begin{equation}
  \label{eq:VGauss}
      V(\mathbf{r}) = V_0e^{-\alpha((x-x_i)^2+z^2)}. 
\end{equation}
can be used to model an impurity in the wire or the effect of a tunable central gate.\cite{soh01}
In this subsection and the ones that follow, the center $(x_i,0)$ of the
potential is chosen to be in the
middle of the wire (i.e.\ $x_i = L/2$ and
$x_i = 0$ in the hard wall and parabolically confined quantum wires
respectively) unless otherwise noted.

\begin{figure}[tbh]
      \includegraphics[width=0.45\textwidth]{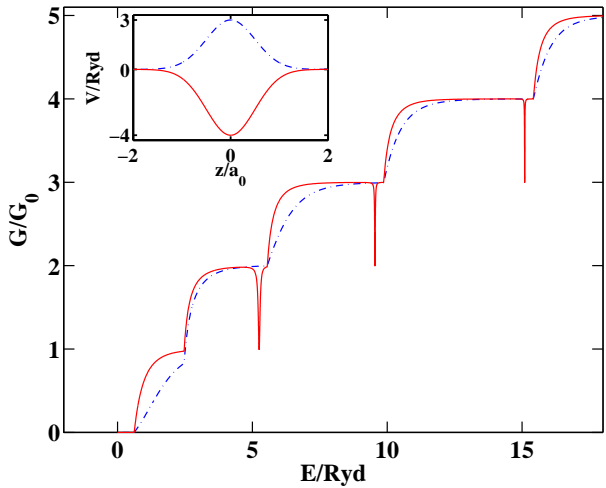}
      \caption{(Color online). Conductance  of a hard wall quantum wire in units of 
               $G_0 = 2e^2/h$ as a function of energy in the presence of a 
               single repulsive (dashed, $V_0 = 3$ Ryd, $\alpha = 2a_0^{-2}$) 
               and attractive (solid, $V_0 = -4$ Ryd, $\alpha = 2a_0^{-2}$) 
               Gaussian scattering potential. The inset shows the scattering 
               potentials in a cross section along the middle of the wire
               ($x=L/2$). The total number of modes is $N=8$. The width of the wire is $L=4a_0$.}
\label{fig:SingleGaussian}
\end{figure}
The conductance of a hard wall quantum wire in the presence of such a
Gaussian scatterer is shown in Fig.~\ref{fig:SingleGaussian}. In the case of a 
repulsive potential, the well known steps in the conductance are little
smeared out, similar to what is seen in experiments.\cite{WHB+88} 
One should remember though that our calculations are at zero temperature, 
so there is no temperature smearing. In experiments, the finite temperature 
can be a significant source of smearing of the steps. Additionally, 
in the case of an attractive potential, there are dips in the conductance 
right before the onset of the next mode. The dips can be understood from a 
simple coupled mode model to be due to backscattering by a quasi-bound 
state originating from an evanescent mode in the next energy subband.\cite{GL93,Bagwell90} 
A more general explanation in terms of the symmetries of the scattering 
matrix also exist.~\cite{NS94} This effect is a multimode effect that
disappears if evanescent modes are not included in the calculations. 
Interestingly, there is no dip in the first subband. This can also be 
understood from the coupled mode model, in which coupling between the 
first propagating mode and the quasi-bound state in the first evanescent 
mode is given by $V_{12}(z)$. Due to symmetry this matrix element
is zero, the potential being even and the wave functions for the 
first and second transverse modes being even and odd respectively, hence no dip.

\begin{figure}[tbh]
      \includegraphics[width=0.45\textwidth]{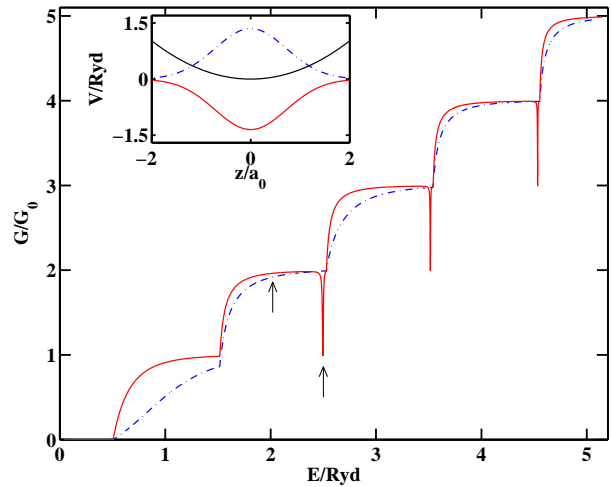}
      \caption{(Color online). Conductance  of a parabolic quantum wire in units of 
               $G_0 = 2e^2/h$ as a function of energy in the presence 
               of a single repulsive (dashed, $V_0 = 1.35$ Ryd, 
               $\alpha = 0.96a_0^{-2}$) and attractive (solid, $V_0 = -1.35$
               Ryd, $\alpha = 0.96a_0^{-2}$) Gaussian scattering potential. 
               The inset shows the scattering potentials in a cross section 
               along the middle of the wire ($x=0$) and the parabolic
               confinement potential ($\hbar\omega = 1.01$ Ryd). 
               The total number of modes is $N=9$. The arrows point at the the
               energies at which the probability density is calculated
               in Figs.~\ref{fig:DensAttractiveNormal} and~\ref{fig:DensAttractiveDip}.}
\label{fig:SingleGaussianParabolic}
\end{figure}
The conductance of a parabolic wall quantum wire in 
the presence of a single Gaussian scattering potential is shown in 
Fig.~\ref{fig:SingleGaussianParabolic}.
Due to different dependence of the transverse energy levels on the subband
number (linear and square) the length of the plateaus increases in the case of 
hard wall potential while being of constant length for the
parabolic wires. Besides this, there is no qualitative difference between the 
two confinement potentials.

The conductance of the hard wall quantum wires has been obtained by solving 
for the $T$-matrix through Eq.~\eqref{eq:TopLSeig2}. The conductance of the 
parabolic wall quantum wire has both been obtained by the $T$-matrix approach 
and by solving the real space LS Eq.~\eqref{eq:LSZsol}. 
In the latter case the real space wave function is a part of the solution but
has to be calculated separately in the
$T$-matrix case (cf.\ App.~\ref{sec:appendixWave}). Visualizing the
wavefunctions can aid in discussing the scattering
processes. Therefore, we now visually examine the system of the parabolically 
confined wire at the energies marked by an arrow in 
Fig.~\ref{fig:SingleGaussianParabolic}, making the physics 
of the process more clear. 
\begin{figure}[tbh]
      \includegraphics[width=0.45\textwidth]{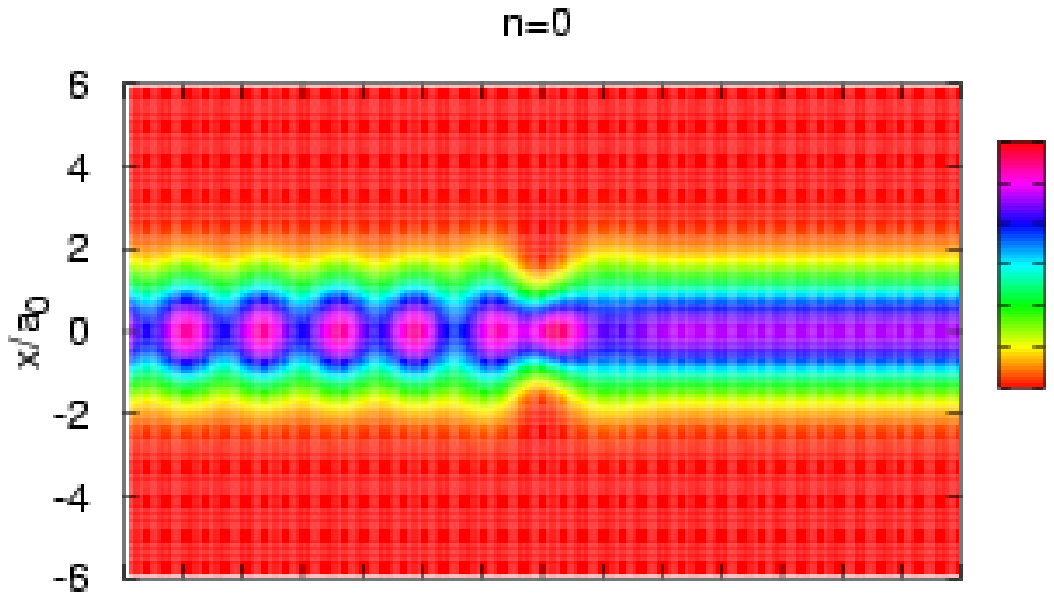}\\
      \includegraphics[width=0.45\textwidth]{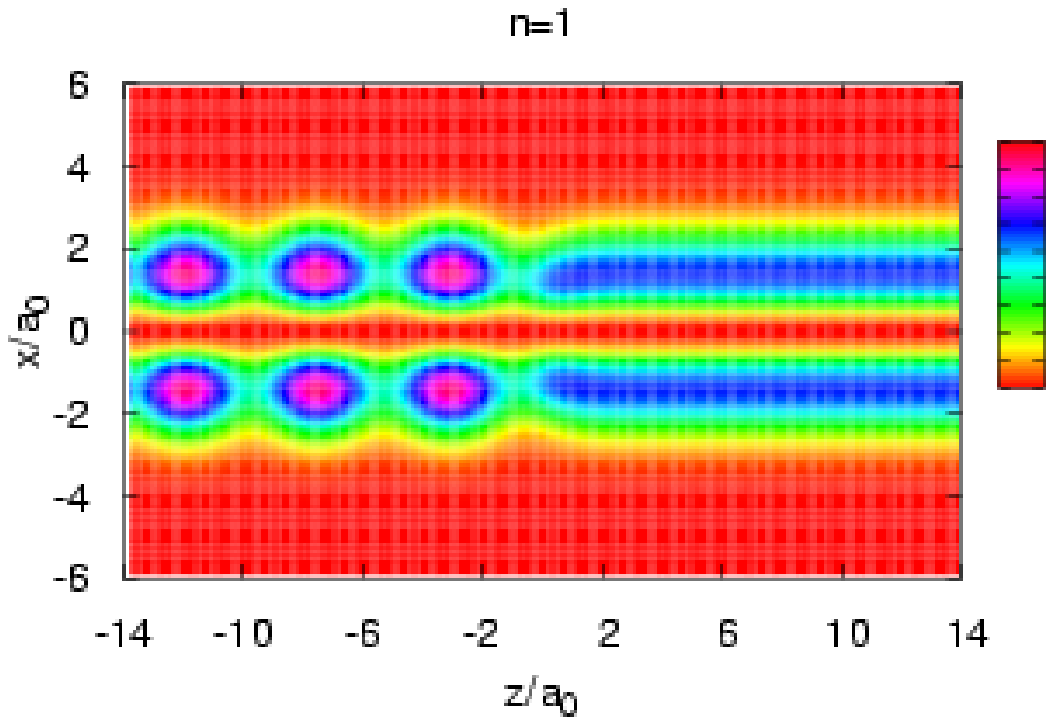}
      \caption{(Color online). The probability density of the scattering states $\psi_{nE}^+$ 
               in the parabolic quantum wire in the presence of the single 
               attractive Gaussian scattering potential of 
               Fig.~\ref{fig:SingleGaussianParabolic}. The total energy of
               the incident particle is $E = 2.02$ Ryd 
               (cf.\ the left arrow in Fig.~\ref{fig:SingleGaussianParabolic}). 
               The incoming wave is in mode $n=0$ (above) and $n=1$ (below).}
\label{fig:DensAttractiveNormal}
\end{figure}

In Fig.~\ref{fig:DensAttractiveNormal} we have plotted the probability 
density $|\psi_{nE}^+(\mathbf{r})|^2$ of the scattering state for incoming waves with $n = 0$ 
and $n = 1$, at the energy indicated  by the left arrow in Fig.~\ref{fig:SingleGaussianParabolic}. 

The incoming wave is partly reflected and partly transmitted. 
To the left of the scattering center the incoming part and the reflected 
part interfere to form the beating pattern seen. To the right, only the 
transmitted part exists and the probability density is independent of $z$. 
It is interesting to note
that in general the transmitted part in the second subband is a linear 
combination of the first two modes. The two modes should then interfere to 
make a beating pattern also on the right hand side. However, as already 
mentioned, the coupling between the first two modes is zero due to symmetry. 
Therefore an incident electron in the first mode cannot be transmitted
into the second mode. Examining a similar picture in the third subband reveals 
interference between mode $n=0$ and mode $n=2$ in the right hand side, 
in agreement with the above. The effect of the normal modes can also be 
seen in  the nodal structure in the transverse direction. 
The second mode, $n=1$, has a node in the middle of the wire while the 
first mode, $n=0$, is nodeless. 

\begin{figure}[tbh]
      \includegraphics[width=0.45\textwidth]{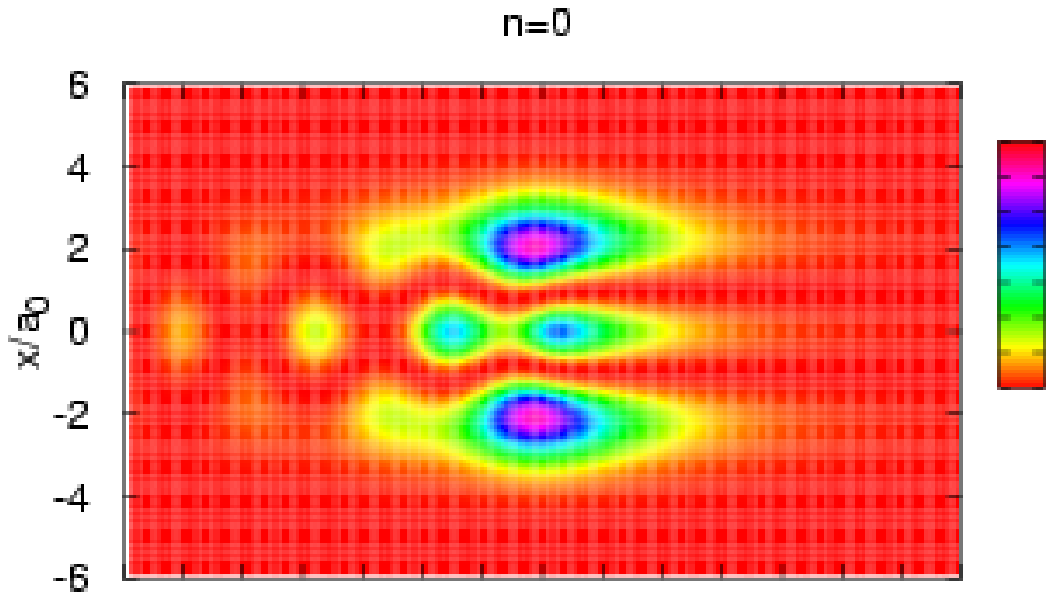}\\
      \includegraphics[width=0.45\textwidth]{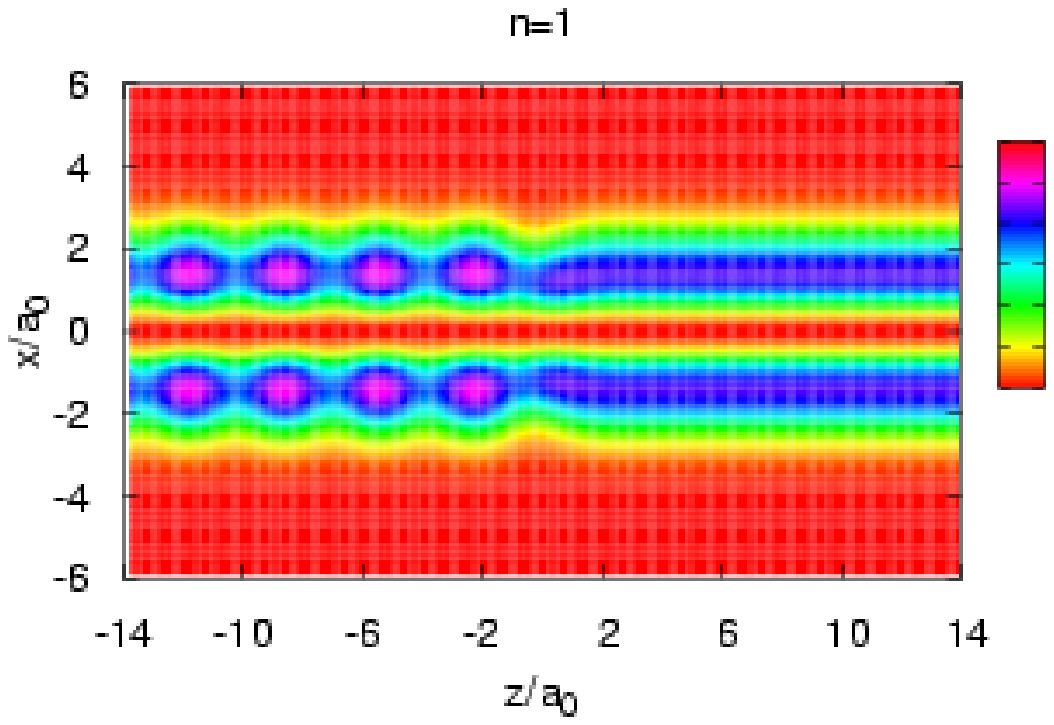}
      \caption{(Color online). The probability density of the scattering states $\psi_{nE}^+$ 
               in the parabolic quantum wire in the presence of the single 
               attractive Gaussian scattering potential of
               Fig.~\ref{fig:SingleGaussianParabolic}. The total energy of
               the incident particle is $E = 2.50$ Ryd, coinciding with a dip in
               the conductance (cf.\ the right arrow in
               Fig.~\ref{fig:SingleGaussianParabolic}). 
               The incoming wave is in mode $n=0$ (above) and 
               $n=1$ (below).}
\label{fig:DensAttractiveDip}
\end{figure}
More interesting things are happening in the dip in the second subband. The upper panel of
Fig.~\ref{fig:DensAttractiveDip} shows the probability density of the scattering state 
obtained when the incident wave is in mode $n=0$.
This mode is resonantly backscattered due to 
coupling to an evanescent state in mode $n=2$. Mode $n=1$ is unaffected 
because of symmetry blocking (lower panel of Fig.~\ref{fig:DensAttractiveDip}).  The  
evanescent state with the corresponding nodal structure is clearly seen in the total wavefunction as a
localized state
around the scattering potential. A beating pattern to the left of the
scattering center, similar to the one seen in
Fig.~\ref{fig:DensAttractiveNormal}, due to the interference of the incoming 
and reflected wave is faintly seen. On the other side of the
potential, however, no outgoing wave is seen in agreement with resonant 
backscattering. An incident wave in mode $n=1$ can due to symmetry not couple 
to the evanescent state. A scattering state with such an incident wave is thus 
expected to be qualitatively the same as for an energy away from the dip in 
the same subband, as can be seen in the lower panel of 
Fig.~\ref{fig:DensAttractiveDip}. The main difference between the scattering 
states with incident wave in mode $n=1$ in Fig.~\ref{fig:DensAttractiveNormal} and
Fig.~\ref{fig:DensAttractiveDip} is that the latter is at higher energy. 
The wave vector is thus higher, resulting in higher frequency in the
oscillations of the probability density, The same is true when comparing the different 
frequencies of the oscillations in the two cases of 
Fig.~\ref{fig:DensAttractiveNormal}. In the $n=0$ case, a smaller part of the 
energy is in the transverse mode thus increasing the wave vector of the wave 
in the propagating direction.

Both of these effects, the steps and the dip, are well known in the
literature. However, in most calculations the
scattering potential is of zero-range delta function 
type.\cite{Bagwell90,VV2002,CM03} Extended potentials are rare, 
an example being the rectangle potential of Bagwell.\cite{Bagwell90} 
An interesting difference can be seen in the total number
of modes needed in the calculations. In our case, 8 modes give results 
that change insignificantly if more modes are added.
In the delta function case, up to 100 modes are needed, a clear 
signature of the singular nature of the delta function potential.

\subsection{A Quantum Dot Embedded in the Wire}
\begin{figure}[tbh]
      \includegraphics[width=0.45\textwidth]{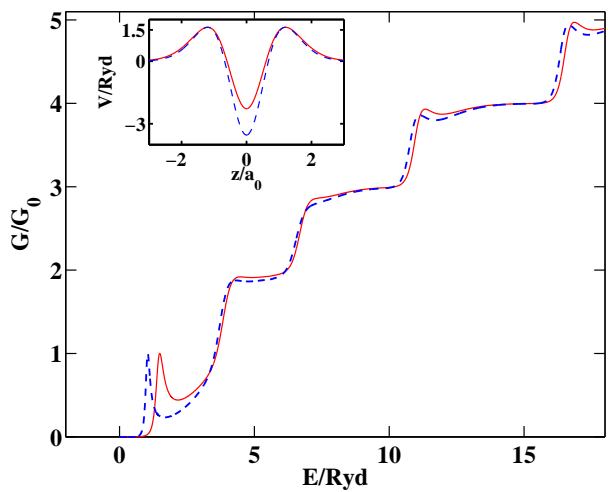}
      \caption{(Color online). Conductance  of a hard wall quantum wire in units of 
               $G_0 = 2e^2/h$ as a function of energy in the presence
               of a double Gaussian scattering potential with a varying 
               depth of the well. The inset shows the scattering
               potentials, whose parameters in the solid (dashed)
               case are $V_1 = 5.21$ ($6.16$) Ryd, $V_2= 7.49$ ($9.70$) 
               Ryd, $\alpha_1 = 0.52$ ($0.60$) $a_0^{-2}$ and $\alpha_2 = 1.52$ ($1.60$)
               $a_0^{-2}$, in the cross section $x=L/2$. The total number of modes
               is $N=8$. The width of the wire is $L=4a_0$.}
\label{fig:DoubleGaussian}
\end{figure}
The double Gaussian
\begin{equation}
  \label{eq:VDoubleGauss}
      V(\mathbf{r}) = V_1e^{-\alpha_1 r^2} - V_2e^{-\alpha_2 r^2} \quad \alpha_2 > \alpha_1, 
\end{equation}
can be used to model a quantum dot embedded in the wire. 
The conductance of a hard wall quantum wire with
such a quantum dot is given in Fig.~\ref{fig:DoubleGaussian}.
Compared to the single impurity, a new effect appears. 
An enhanced conductance in the first subband along 
with similar weaker structures in higher bands. This is the well known 
effect of resonant tunneling. A remnant of a bound state in the well induces 
a resonance in the transmission, hence the resonance in the conductance. 
\begin{figure}[tbh]
      \includegraphics[width=0.45\textwidth]{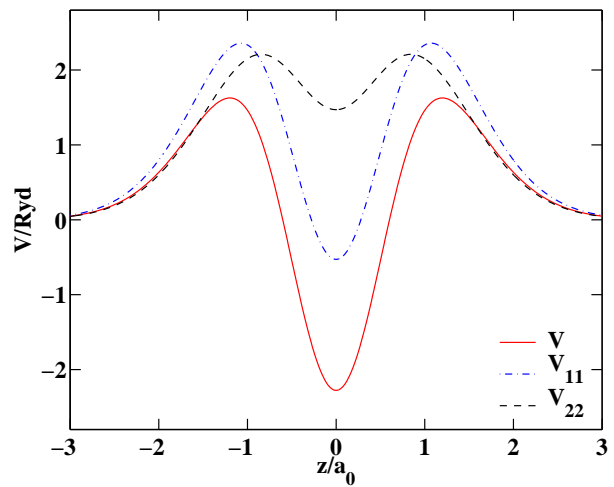}
      \caption{(Color online). The solid line represents the double Gaussian potential of 
               Fig.~\ref{fig:DoubleGaussian} along with its first two
               effective potentials.}
\label{fig:Veffective}
\end{figure}
Examining the coupled mode equation~\eqref{eq:SchrodingerZ}, one notices that, 
ignoring the coupling between the modes, the transmission is governed by the 
effective potentials $V_{nn}(z)$. In Fig.~\ref{fig:Veffective} we have plotted 
the double Gaussian potential along with its first two effective potentials. 
We know from resonant tunneling in 1D, that the
higher the probability for the electron to tunnel out of the well, 
the broader the resonance in the transmission. The difference in the sharpness
of the tunneling resonances in different subbands can therefore be attributed 
to the different shapes of the effective potential. 
\begin{figure}[tbh]
      \includegraphics[width=0.45\textwidth]{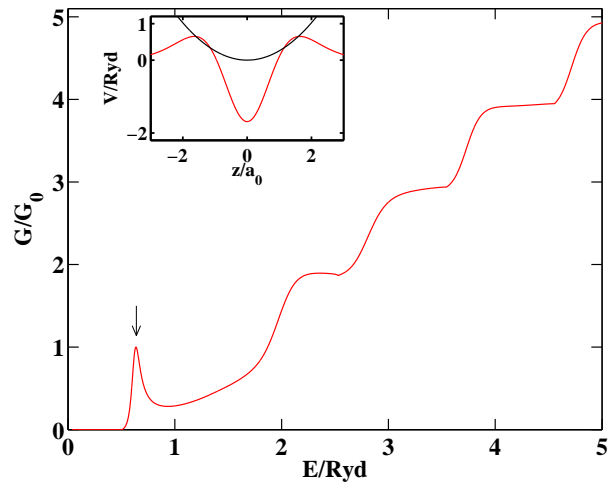}
      \caption{(Color online). Conductance  of a parabolic wall quantum wire in units of 
               $G_0 = 2e^2/h$ as a function of energy in the presence
               of a double Gaussian scattering potential. The inset shows 
               the scattering potential, whose parameters are $V_1
               = 2.02$ Ryd, $V_2= 3.71$ Ryd, $\alpha_1 = 0.29$ 
               $a_0^{-2}$ and $\alpha_2 = 0.96$ $a_0^{-2}$, in the cross 
               section $x=0$. The parabolic confinement potential 
               ($\hbar\omega = 1.01$ Ryd) is also shown. The total 
               number of modes is $N=9$. The arrow points at the value of the
               energy at which the probability density in
               Fig.~\ref{fig:DensAttractiveResonance} is calculated.}
\label{fig:DoubleGaussianParabolic}
\end{figure}

The conductance through a quantum dot in a parabolic wire in 
Fig.~\ref{fig:DoubleGaussianParabolic} has the
same pronounced resonance in the first subband but the resonances in the
higher subbands are even more
suppressed than in the hard wall wire.

\begin{figure}[tbh]
      \includegraphics[width=0.45\textwidth]{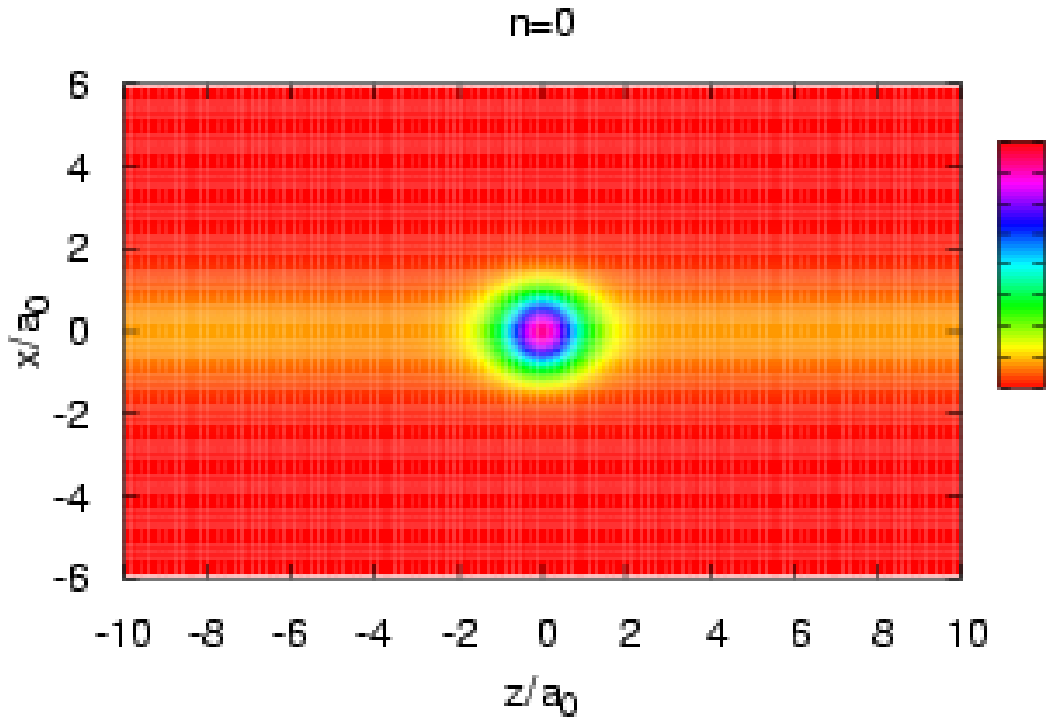}
      \caption{(Color online). The probability density of the scattering states 
               $\psi_{nE}^+$ in the parabolic quantum wire in the
               presence of the double Gaussian scattering potential 
               of Fig.~\ref{fig:DoubleGaussianParabolic}. The total energy of
               the incident particle is $E = 0.64$ Ryd, coinciding with a 
               resonance in the conductance (cf.\ the arrow in 
               Fig.~\ref{fig:DoubleGaussianParabolic}). 
               The incoming wave is in mode $n=0$.}
\label{fig:DensAttractiveResonance}
\end{figure}

The resonant tunneling is 
through the remnants of a bound state in the well. 
In contrast to the resonant backscattering, this quasi-bound state belongs to the
propagating subband, not the next evanescent mode. The resonance is thus not a 
multimode phenomenon as the dip certainly is. We examine the probability density in the
most pronounced resonance of Fig.~\ref{fig:DoubleGaussianParabolic}, that is
the one marked by an arrow. There is only one propagating mode 
and thus the only scattering state is the one with an incident wave in mode $n=0$; 
the probability density of which is shown if Fig.~\ref{fig:DensAttractiveResonance}. 
The quasi-bound state in the middle of the wire is clearly seen. 
The absence of beating on both sides of the scattering center indicates
that there is no reflected wave, i.e.\ the incoming wave is perfectly
transmitted. This supports very well the suggestion that the resonance is 
due to resonant tunneling through a remnant of a bound state in the well.

\subsection{Shape Effects of the Scattering Potential}
\begin{figure}[tbh]
      \includegraphics[width=0.45\textwidth]{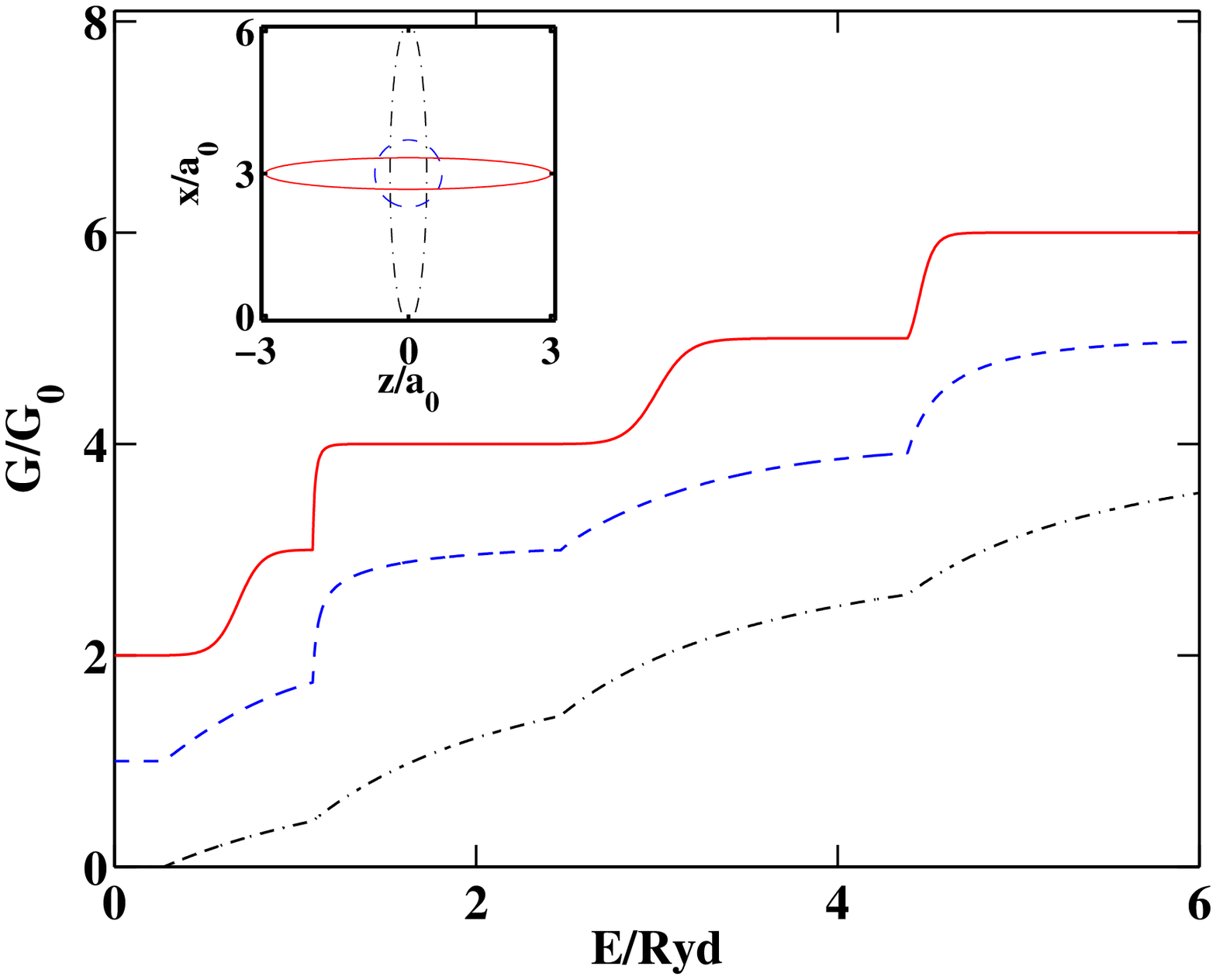}\\
      \includegraphics[width=0.45\textwidth]{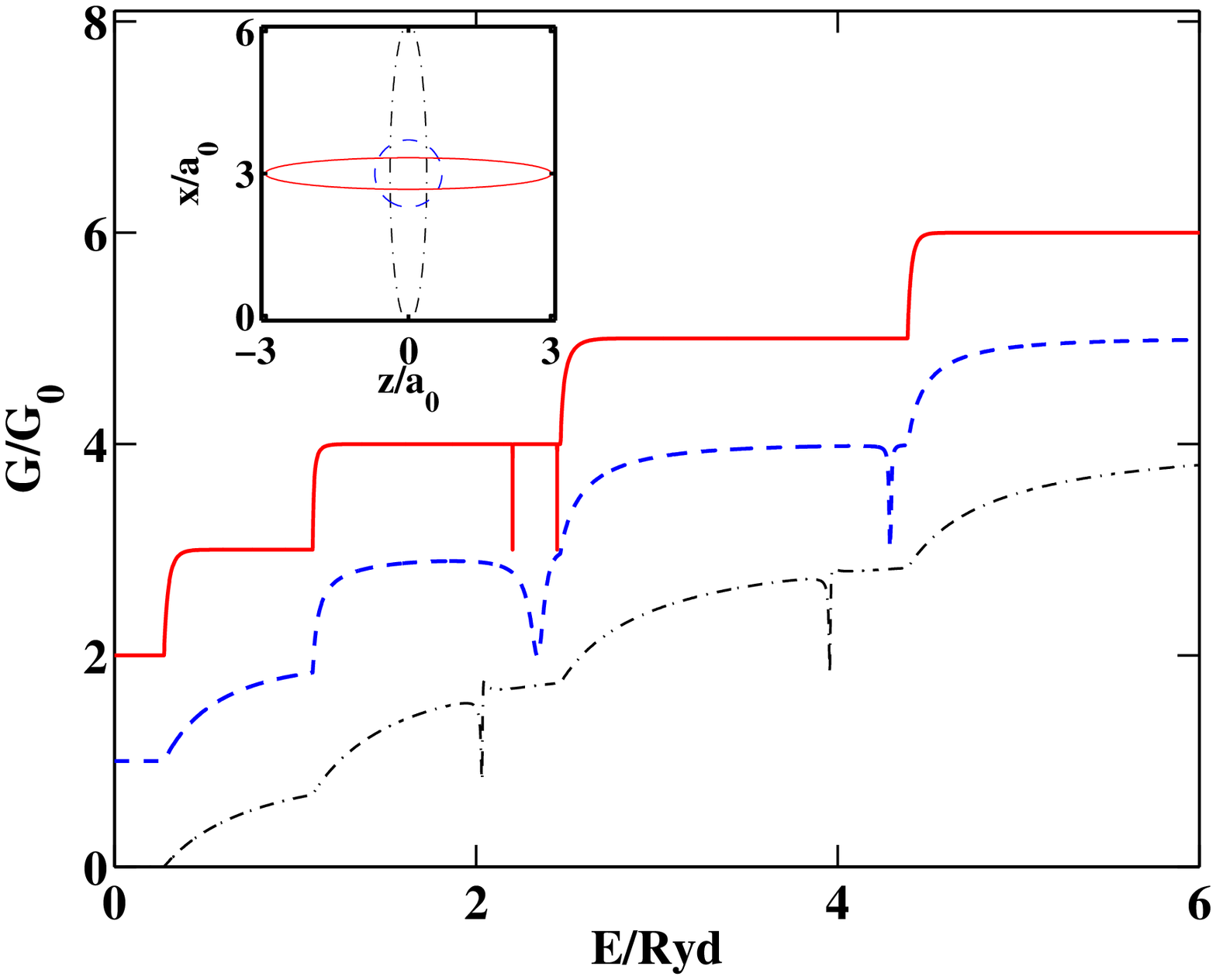}
      \caption{(Color online). Conductance of a hard wall quantum wire in units of $G_0 =
              2e^2/h$, as a function of energy. A single contour of the scattering
              potential, an asymmetric single Gaussian, is shown in the inset. 
              The contour is chosen to be at the value of
              $V_0e^{-1} = 0.368V_0$. In the upper panel the potentials are
              repulsive ($V_0 = 3$ Ryd) but attractive ($V_0 = -3$ Ryd) in the
              lower panel. In both cases the circularly symmetric potential
              has $\alpha = 2$ $a_0^{-2}$, the transverse one has $\alpha_x =
              0.10$ $a_0^{-2}$ and $\alpha_z = 6.73$ $a_0^{-2}$, and the
              longitudinal has $\alpha_x = 9.10$ $a_0^{-2}$ and $\alpha_z =
              0.11$ $a_0^{-2}$. The parameters are chosen such that the 
              volume between the potential and the zero energy plane are
              equal. The width of the wire is $L = 6$ $a_0$ and the total number of
              modes is $N=8$. The symmetric and longitudinal potentials have 
              been shifted by $1$ $G_0$ and $2$ $G_0$
              respectively, for clarity.}
\label{fig:Assymetric}
\end{figure}
So far, the scattering potentials under examination have all been circularly
symmetric. Elongated extended scattering potentials, such as the asymmetric
Gaussian 
\begin{equation}
  \label{eq:VGaussAsymmetric}
      V(\mathbf{r}) = V_0e^{-\alpha_x(x-x_i)^2-\alpha_zz^2}, 
\end{equation}
offer more variety. The conductance characteristics for the cases of elongated
longitudinal, elongated transverse, and circularly symmetric
potential profiles are under investigation, as shown in Fig.~\ref{fig:Assymetric}. 
The parameters of the potentials have been chosen such that their volume in the wire is the
same in all cases. This is done so that the strength of the scatterers is
comparable in all cases. We have used both attractive and repulsive potentials.

We begin by discussing the conductance of the repulsive potentials in the
upper panel of Fig.~\ref{fig:Assymetric}. The conductance of the transverse 
barrier is lower than its circularly symmetric counterpart. The conductance of the longitudinal
barrier has a more pronounced exponential growth and is thus lower than the 
conductance of the circularly symmetric barrier at the low energy
part of each subband while being higher in the high energy end. Both the
symmetric and the longitudinal 
barrier are mainly confined to the middle of the wire. 
Modes that have nodes in the middle of the wire,
i.e.\ the second, fourth etc.\ mode, see very little of the potential and go easily through. This is
especially clear in the case of the longitudinal potential were these modes
have almost perfect transmission as soon as they become propagating. 
The transverse barrier is nearly independent of the transverse direction and
affects all the modes equally. 

The lower panel of Fig~\ref{fig:Assymetric} shows the conductance of
attractive potentials. Compared to the
symmetric case, the elongated longitudinal and transverse attractive barriers 
have higher and lower conductance respectively. The dip structure is quite
different. The shape of the dips in the transverse case are more of
the asymmetric Fano form than the more common Breit-Wigner form as in the
longitudinal case.\cite{NS94} In the longitudinal case, we have two dips rather than
the usual single dip. Because of symmetry blocking, there are no dips in the
first band as before. Even though no dips are seen
in the third subband we belive they are there. They are just to narrow to be seen in the energy
resolution of the calculations.

\begin{figure}[tbh]
      \includegraphics[width=0.45\textwidth]{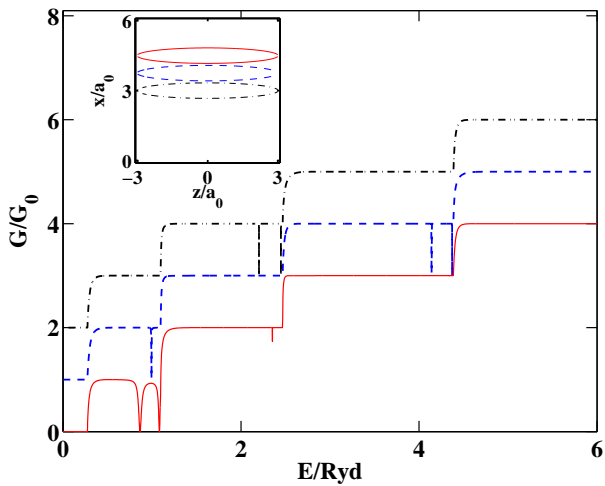}
      \caption{(Color online). Conductance of a hard wall quantum wire in units of $G_0 =
               2e^2/h$, as a function of energy. The scattering
               potential has the same parameters as the attractive 
               potential in Fig.~\ref{fig:Assymetric}, but the center is moved
               from the middle of the wire $x_i = L/2$ to $x_i = 5L/8$ and $x_i =3L/4$. 
               The width of the wire is $L = 6$ $a_0$ and the total
               number of modes is $N=8$ as before. The inset shows the
               $V_0e^{-1}$ contour of the scattering potentials. 
               The potentials in the center of the wire and the one with $x_i
               = 5L/8$ have been shifted by $1$ $G_0$ and $2$ $G_0$
               respectively, for clarity.}
\label{fig:AssymetricCenter}
\end{figure}
The dips are most probably too narrow to be
seen in experiments. By shifting the center of the potential from the middle
of the wire, dips in the first
subband are no longer blocked (cf.\ Fig.~\ref{fig:AssymetricCenter}). When the center of the
longitudinal potential is in $x_i = 3L/4$
we have two dips that are quite pronounced. We will now focus on these two dips.

\begin{figure}[tbh]
      \includegraphics[width=0.45\textwidth]{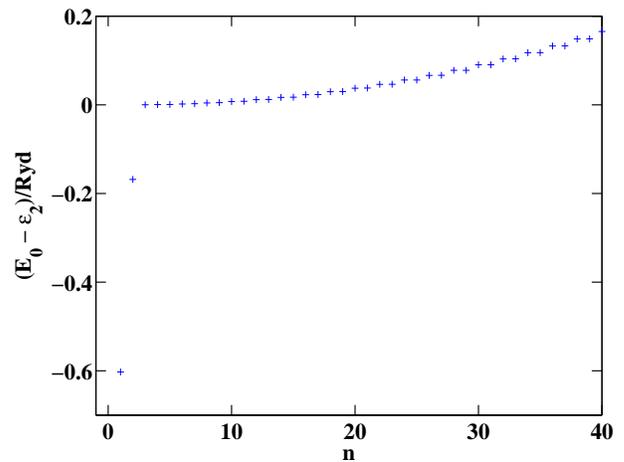}
      \caption{The energy levels of the effective potential $V_{22}(z)$ of the 
               longitudinal attractive potential of
               Fig.~\ref{fig:AssymetricCenter} with $x_i=3L/4$. 
               The levels are obtained in a box of
               length $L_{\text{box}} = 150$ $a_0$.}
\label{fig:QuasiBoundEnergyLevels}
\end{figure}

If we include only one mode, i.e.\ the propagating mode, in the calculations 
these dips disappear. Both dips reappear by adding
just the first evanescent mode. The higher evanescent modes change the conductance curve
only quantitatively, shifting the dips and affecting their width. We therefore deduce that both dips
can be explained by a coupled mode model restricted to two modes. 
This kind of model as already been put
forth to explain the single dip in the presence of a single
impurity.\cite{GL93,NS94} N{\"o}ckel and Stone
noted that a quasi-bound state $\Phi$, satisfying
\begin{equation}
\label{eq:QuasiBound}
\left(-\frac{\hbar^2}{2m}\frac{d^2}{dz^2} + V_{22}(z)\right) \Phi(z) = (E_0 - \varepsilon_2)\Phi(z)
\end{equation}
appears from the evanescent mode. They note that there is no need for the dip
to be close to the subband
edge\footnote{see footnote in Ref.~\onlinecite{NS94}}. Even though they do not mention it, it is apparent
from their formalism that there is also nothing that prevents more than one
quasi-bound state to be present, as
long as there are more than one solution to Eq.~\eqref{eq:QuasiBound} with
$E_0-\varepsilon_2 < 0$. We therefore suggest that the two dips are due to two 
quasi-bound states originating from the first evanescent
mode. To confirm this suggestion we have solved Eq.~\eqref{eq:QuasiBound} 
numerically by putting a large box around the potential and expanding the
bound state $\Phi$ in the plane-wave eigenfunctions of the particle
in a box with periodic boundary conditions. The energy eigenvalues obtained are plotted in
Fig.~\ref{fig:QuasiBoundEnergyLevels}, showing clearly that there are two
bound states, that develop into
quasi-bound states due to coupling to the propagating state.

Depending on the shape and strength of the scattering potential the above
argument suggests that one could
see even more dips in each subband. Furthermore, if the binding energy
fulfills $E_0 - \varepsilon_2 <
-(\varepsilon_2 - \varepsilon_1)$, a true bound state originating from the 
second mode would exist.

\begin{figure}[tbh]
      \includegraphics[width=0.45\textwidth]{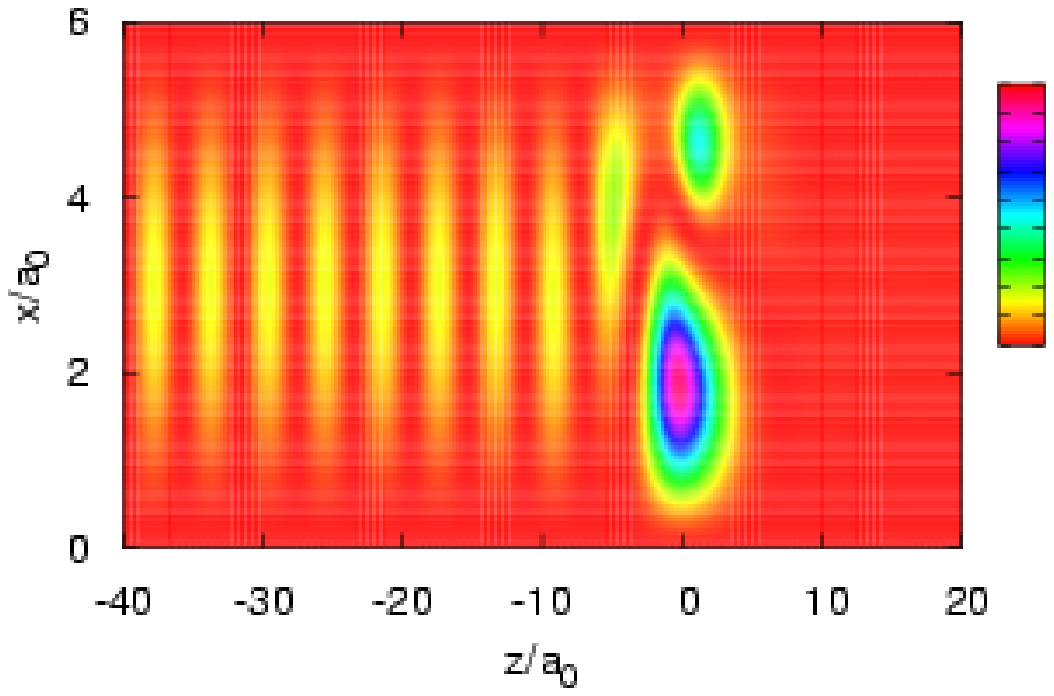}\\
      \includegraphics[width=0.45\textwidth]{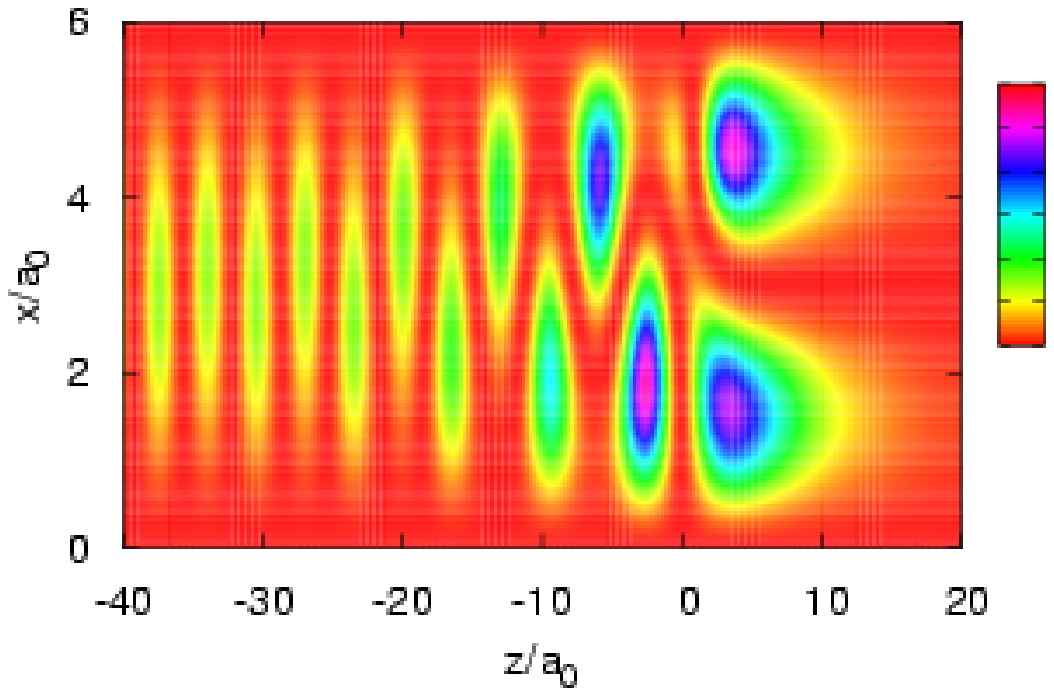}
      \caption{(Color online). The probability density of the scattering states $\psi_{nE}^+$ 
               in the hard wall quantum wire in the presence of the elongated
               longitudinal Gaussian potential shown in the inset of
               Fig.~\ref{fig:AssymetricCenter} shifted to $x_i = 3L/4$. 
               The incident wave is in mode $n=1$ and has
               total energy $E = 0.865$ ($1.084$) Ryd in the upper (lower)
               panel. These energies correspond to the two
               dips in the conductance of Fig.~\eqref{fig:AssymetricCenter}.}
               \label{fig:DensDoubleDip}
\end{figure}
To confirm our suggested mechanism, in Fig.~\ref{fig:DensDoubleDip}, we examine
the probability density of the
scattering states at the energies corresponding to the two dips. 
The symmetries of the quasi-bound states are as expected. 
The one lower in energy is nodeless in the $z$ direction while
the other one has a single node. Both have the symmetry of the transverse mode $n=2$.

We note in passing that the width of the scattering potential can be estimated as the width of the
$V_0e^{-1}$ contour. The elongated potential in
Fig~\ref{fig:AssymetricCenter} has the width $0.33$ $a_0$ in the transverse
direction and $3.02$ $a_0$ in the propagating
$z$ direction. We observe that the quasi-bound states are much more extended
than one would expect of a true bound state
in the potential. They also have probability density on both sides of the wire. 
The above indicates that we are looking at the two
quasi-bound states from mode $n=2$.

\section{\label{sec:summary}Summary and discussion}
In this work we have studied the coherent quantum transport in the 
presence of various realistic Gaussian-type scattering potentials using 
the Lippmann-Schwinger $T$-matrix approach.  We have calculated the 
conductance and the spatial distribution of the electron probability and their
dependence on the shape of the scattering potentials.
The formalism is quite general and can be applied to quantum wires modeled with 
different confinement potentials, and in the presence of a general extended 
scattering potential. We have used both hard wall
confinement and parabolically confined quantum wires. 
The scattering potential has been modeled by Gaussian
potential, using single and double, symmetric and asymmetric Gaussians. 
By a combination of these we have modeled a single impurity, a quantum dot
embedded in the wire and studied the shape effects of the
scattering potential.  Repulsive impurity smooths out the exact quantization
of the conductance while
attractive impurity shows the well known dips in the conductance. 
A resonant transmission appears in the
conductance through the quantum dot as the energy of the incident electron
coincides with an energy level in the dot. A longitudinal barrier affects 
modes of different symmetry differently while a transverse barrier affects all 
the modes in the same way. In an attractive longitudinal potential two dips
can appear in the same subband. These two dips appear due to coupling of
a propagating mode to two quasi-bound states that both originate in the next 
evanescent mode. 

Our calculations demonstrate the power and flexibility of the current
approach. Being able to calculate the conductance of a quantum wire in 
the presence of a general extended scattering potential allows for a multitude of
systems to be studied. Our Gaussian potentials are rather simple but by 
appropriately choosing the scattering potential many
different experimental setups can be modeled.

\appendix
\section{\label{sec:appendix}Solving the $T$-matrix Integral Equation}
When solving the LS equation for the $T$-matrix, Eq.~\eqref{eq:TopLSeig2},
numerically, one has to treat the singularities of the Green's
function with care. By using the fact that 
\begin{equation}
\label{eq:GreensPrincipal}
      \frac{1}{k_l^2-q^2+i\eta} = \frac{{\cal P}}{k_l^2-q^2} - 
      \frac{i\pi}{2k_l}\left( \delta(q-k_l) + \delta(q+k_l) \right),
\end{equation}
the integral on the right hand side can be transformed into two algebraic
terms plus a principal value integral (${\cal P}$ denotes a principal value)
\begin{equation}
\label{eq:PrincipalValue}
      {\cal P}\int_{-\infty}^\infty dq\,\frac{V_{ml}(k,q)T_{ln}(q,k_n)}{k_l^2-q^2}.
\end{equation}
Since
\begin{equation}
\label{eq:ZeroIntegral}
      {\cal P}\int_{0}^{\infty} dq\, \frac{1}{k_l^2-q^2} = 0,
\end{equation}
it is possible to remove the singularity of the integrand by a subtraction of a 
zero~\cite{HT70,RLandau}. For a general function $f$ we have (assuming $k>0$)
\begin{equation}
\label{eq:ErasingSingularity}
\begin{split}
      {\cal P}&\int_{-\infty}^\infty dq\,\frac{f(q)}{k_l^2-q^2} = {\cal P}\int_{0}^\infty dq\,
      \frac{f(-q)+f(q)}{k_l^2-q^2}=  \\
      &\int_{0}^\infty dq\,\frac{f(-q)-f(-k_l)}{k_l^2-q^2} + \int_{0}^\infty dq\,
      \frac{f(q)-f(k_l)}{k_l^2-q^2}.
\end{split}
\end{equation}
The value of the integrand at the singularity can, by use of
l'H$\hat{\text{o}}$pital's rule, be seen to be
$\mp f'(\pm k)/2k$. Assuming that $f'$ is nonsingular at $\pm k$, the principal 
value can therefore be skipped as already done. Using this prescription and 
numerical Gaussian integration for the remaining integrals, 
the integral LS equation for the $T$-matrix is transformed into a system of 
linear equations that can be solved by standard linear algebra
methods.\cite{JHBMSThesis} 

The Green's function for the evanescent modes is non singular since $k_l^2 <
0$. The integral in Eq.~\eqref{eq:TopLSeig2} for evanescent modes is therefor
done with numerical Gaussian integration without the above special treatment
for singularities.

\section{\label{sec:appendixWave}Scattering states in terms of the T-matrix}
By a spectral representation of the mode Green's function, Eq.~\eqref{eq:Green_nE},
\begin{equation}
  \label{eq:Green_1D_spectral}
  G_{mE}^0(z,z') = \frac{1}{2\pi}\int_{-\infty}^{+\infty} 
  dp \frac{e^{ip(z-z')}}{k_m^2 - p^2},
\end{equation}
the wave function in Eq.~\eqref{eq:LSZsol} can be rewritten as
\begin{equation}
  \label{eq:phi_m}
  \begin{split}
    \phi_{mE}^n(z) 
    &= \phi_{mE}^{n0}(z) + \frac{m}{\pi\hbar^2}
    \sum_{m'} \int_{-\infty}^{+\infty} dp \frac{e^{ipz}}{k_m^2-p^2}\\
    &\times\iint dx'dz' e^{-ipz'}\chi_m^*(x')V(\rb')
    \phi_{m'E}^n(z')\chi_{m'}(x') ,
  \end{split}
\end{equation}
which further yields
\begin{equation}
  \label{eq:phi_m2}
  \begin{split}
    \phi_{mE}^n(z) 
    &= \phi_{mE}^{n0}(z) + \frac{m}{\pi\hbar^2}
    \int_{-\infty}^{+\infty} dp \frac{\sqrt{|p|}e^{ipz}}{k_m^2-p^2}\\
    &\times\iint dx'dz' \frac{e^{-ipz'}}{\sqrt{|p|}}
    \chi_m^*(x')V(\rb')\Psi_{nE}^+(\rb'),
  \end{split}
\end{equation}
where we have used the expansion in Eq.~\eqref{eq:SSexpansion}. By the definition of the
T-matrix, Eq.~\eqref{eq:Tmatrix3D}, we can thus rewrite Eq.~\eqref{eq:phi_m2} as
\begin{equation}
  \label{eq:phi_T}
  \begin{split}
  \phi_{mE}^n(z)
    = \phi_{mE}^{n0}(z) + \frac{m}{\pi\hbar^2}
    \int_{-\infty}^{+\infty} 
    dp \frac{\sqrt{|p|}e^{ipz}}{k_m^2-p^2}T_{mn}(p,k_n). 
  \end{split}
\end{equation}
We have thus expressed the wave function in terms of the elements of the
T-matrix. By Eq.~\eqref{eq:phi_T} and the expansion in Eq.~\eqref{eq:SSexpansion} we
are able to determine the scattering states $\Psi_{nE}^+$ for all $\rb$. 
In the case of propagating modes, the singularities of the mode Green's
function are treated the same way as in App.~\ref{sec:appendix}. For evanescent modes there
are no singularities and we furthermore have that the incoming wave,
$\phi_{mE}^{n0}(z)$, is zero.

\begin{acknowledgments}
This work was partly funded by the Icelandic Natural Science Foundation,
the University of Iceland Research Fund, and the National Science Council
of Taiwan. C.~S.~T. acknowledges the hospitality 
of University of Iceland where this work was initiated.
\end{acknowledgments}

\end{document}